\documentclass[12pt,preprint,tighten,flushrt]{aastex}

\usepackage{epsfig}
\usepackage{colordvi}
\usepackage{html}

\newcommand{\rtp}[1]{\ensuremath{^{#1}}}

\newcommand{\pmt}{\ensuremath{\pm}}

\newcommand{\mas}{\ensuremath{\mbox{mas}}}
\newcommand{\muas}{\ensuremath{\mu\mbox{as}}}

\newcommand{\ph} [1]{\phantom{#1}}

\newcommand{\HIPPARCOS}{{\em HIPPARCOS}}

\newcommand{\SIM}      {{\em SIM}}
\newcommand{\GAIA}     {{\em GAIA}}

\newcommand{\TYCHOT}   {{\em TYCHO-2  }}

\newcommand{\ARIHIP}   {{\em ARIHIP   }}

\newcommand{\eg}{{\it e.g.},\ }
\newcommand{\etal}{{\it et al.}}

\newcommand{\snROv}[1]{\renewcommand{\baselinestretch}{#1}\begin{normalsize}}
\newcommand{\enROv}{\end{normalsize}\renewcommand{\baselinestretch}{1.0}}

\newcommand{\ssROv}[1]{\renewcommand{\baselinestretch}{#1}\begin{small}}
\newcommand{\esROv}{\end{small}\renewcommand{\baselinestretch}{1.0}}

\begin{document}

\title{Finding Solar System Analogs With SIM and HIPPARCOS\\
       \vspace*{1em}
       A White Paper for the Exo Planet Task Force}

\author{\vspace*{1cm}
        {\Large March 2007}\\
        \vspace*{1cm}
        Rob P. Olling\rtp{1}}

\affil{
\rtp{1}University of Maryland at College Park\\
       email: olling@astro.umd.edu
}
\section*{}

\maketitle

\clearpage

\setcounter{page}{1}

 \section*{Abstract}
\label{sec:Abstract}
\vspace*{-0.9em}

The astrometric signature imposed by a planet on its primary increases
substantially towards longer periods ($\propto \, P^{2/3}$), so that
long-period planets can be more easily detected, in principle. For
example, a one Solar-mass (M$_{\odot}$) star would be pulled by
roughly 1 milli-arcsec (mas) by a one Jupiter-mass (M$_J$) planet with
a period of one-hundred years at a distance of 20 pc [cf.
eqn.~(\ref{eqn:a_o_M_J}) below]. Such position accuracies can now be
obtained with both ground-based and space-based telescopes. The
difficulty was that it often takes many decades before a detectable
position shift will occur. However, by the time the next generation of
astrometric missions such as \SIM\footnote{
\label{foot:sim_book}
%
http://planetquest.jpl.nasa.gov/documents/WhitePaper05ver18\_final.pdf
} [\eg \citep{SIM05}] will be taking data, several decades will have
past since the first astrometric mission,
\HIPPARCOS\ \citep{ESA97}. \\
\hspace*{1.4em}
Here we propose to use a new astrometric method that employs a future,
highly accurate \SIM\ Quick-Look (SQL) survey and \HIPPARCOS\ data
taken twenty years prior. Using a conservative position error for
\SIM\ of 10 \muas, this method enables the detection and
characterization of ``Solar-system analogs'' (SOSAs) with periods up
to 80 (165) years for 1 (10) M$_J$ companions. Employing the standard
\SIM\ error of 4 \muas, this period range is extended by a factor of
two to four. We might expect the PDF to turn over in this period
regime.\\
\hspace*{1.4em}
Because many tens of thousands nearby stars can be surveyed this way
for a modest expenditure of \SIM\ time and SOSAs may be quite
abundant, we expect to find many hundreds of extra-solar planets with
long-period orbits. Such a data set would nicely complement the
short-period systems found by the radial-velocity method. Brown dwarfs
and low-mass stellar companions can be found and characterized if
their periods are shorter than about 500 years.  This data set will
provide invaluable constraints on models of planet formation, as well
as a database for systems where the location of the giant planets
allow for the formation of low-mass planets in the habitable zone.

\vspace*{-2.5em}
  \section{Introduction}
\label{sec:Introduction}
\vspace*{-0.9em}
This white paper is based on very recent work summarized in a review
paper on \SIM-science \citep{SIM_PASP}, while we present more details
elsewhere \citep{SOSAs}.\\
\hspace*{1.4em}
Our current knowledge of the demography of extra-solar planetary
systems is mostly a result of long-term radial velocity (RV)
monitoring of nearby (mostly) FGK main-sequence (MS) stars. The
results are spectacular with over 200 suspected planets in $\ga$171
systems \citep{E_ESGP}, with most of the planets in short-period
orbits. Only 10\% of the observed planets have periods exceeding 5
years, while just one (0.5\%) has a period slightly longer than the
period of Jupiter (11.9 yr).  We use the PDF of extra-solar giant
planets (ESGPs) of \citet{TT2002} but scaled-up by a factor 1.6 to
account for current understanding the ESGP frequency [e.g.,
\citet{S2005}].  The updated PDF indicates that the period range
between 5 years and the maximum currently known period should account
for 25\% of the total number of ESGPs rather than the observed
10\%. In fact, the PDF of ESGPs {\em in}creases towards longer periods
($P$) so that systems dominated by long-period planets such as the
Solar system may be quite common, and we will use it to estimate the
frequency of solar system analogs (SOSAs). However, the PDF$_{ESGP}$
{\em has} to turn over at some period to yield a finite integrated
probability. If we {\em had to} guess where the planetary PDF might
turn over, we might pick the period where the stellar PDF turns over,
or about 170 years \citep{DM1991}. \\
\hspace*{1.4em}
We loosely define a SOSA as a system with a (single) planet in the
mass range between Jupiter and Uranus/Neptune ($\sim$0.05 M$_J$) and
with periods between 11.9 years ($P_{Jupiter}$) and 165 years
($P_{Neptune}$). Integrating the PDF$_{ESGP}$ over these ranges, we
find that 12.6\% of systems would be solar-system analogs. If we
consider the group of long-period planets that can be detected
astrometrically, we need to consider more massive systems with masses
($M$) between 1 and 13 M$_J$. We call such systems heavy SOSAs, or
HOSAs. The PDF$_{ESGP}$ predicts that such systems make up about 20\%
of the total number of planetary systems with periods up to 165 years,
and occur around 7.9\% of apparently single stars\footnote{Recent work
[Cumming \etal\ (in preparation) as previewed by \citet{Bea2006}]
indicates that the mass function declines more rapidly towards higher
masses: $dN/dM \propto M^{-1.9}$ rather than $dN/dM \propto M^{-1.1}$
as derived by TT2002. As a result, the {\em number} and relative
frequency of HOSAS would decrease, while their {\em detectability} is
unchanged. Because we deal with detectability, we will use the
re-scaled TT2002 PDF.}.\\
\hspace*{1.4em}
NASA's \SIM\ PlanetQuest can detect extra-solar planets weighing
several times the mass of the Earth [e.g., \citet{Cea2006}].  If
multiple planets exist, their properties can also be determined with
\SIM\ [\eg \citet{SCBL2003,F2006}]. The mission-end astrometric
accuracy of ESA's \GAIA\ astrometric mission is about twenty times
worse than \SIM's, rendering it not very useful for the project
described here.

\vspace*{-2.5em}
  \section{Finding Solar-System Analogs}
\label{sec:Finding_Solar_System_Analogs}
\vspace*{-0.9em}
There are several astrometric methods that can be used to identify
``long-period'' companions of stars. These methods are based on the
fact that the actual proper motion ($\mu$) is non-linear if it
contains a contribution from the reflex motion of the primary being
orbited by a companion. In the ``$\Delta \mu$ method'' a substantial
difference between $\mu$ values from a ``short-term'' catalog such as
\HIPPARCOS\ and those from a ``long-term'' proper motion catalog
such as \TYCHOT\ is indicative of binarity
\citep{WDJLS1999,Wea2000}. \citet{MK2005} find that both the period
and mass can be estimated from the acceleration ($\dot{\mu}$) and the
jerk ($\ddot{\mu}$), but {\em only if} the ``long-term'' proper motion
is known. \citet{KM2003} developed a method that is appropriate for
objects with periods up to twice the mission duration (i.e., up to 10
to 20 years for \SIM).

  \subsection{Past and Future Astrometry}
\label{sec:P3st_and_Future_Astrometry}
\vspace*{-0.9em}
Here we propose a new method to identify and quantify long-period
systems comprising planetary, brown-dwarf (BD) and main-sequence (MS)
companions. Our method uses astrometry of earlier epochs but uses the
positions rather than the proper motions. The other component is a
future, highly accurate astrometric mission such as \SIM.  The idea is
to fit the \SIM\ data for a given star with a simple astrometric
models [e.g., linear, quadratic, etc.], and use this model to {\em
predict} the position of the star at the \HIPPARCOS\ epoch
($\tau_H$). We assume that \SIM\ data will be from 2013.5, leading
to and epoch difference of 22 years. We assume that the \HIPPARCOS\
data are accurate to \pmt 1 \mas. However, this accuracy can be
improved upon in the future by more careful modeling of the systematic
effects, or by using the much improved \GAIA\ reference frame to
define the frame at the \HIPPARCOS\ epoch. The former method has been
applied recently by \citet{vLF2005} who reduced the \HIPPARCOS\ errors
by almost a factor of three, while the latter has been used in the
construction of the \TYCHOT\ catalog \citep{Tycho2}.

\vspace*{-2.5em}
  \section{Model Details}
\label{sec:Model_Details}
\vspace*{-0.9em}
We assume circular, face-on orbits and neglect all the details
associated with orbit fitting. The work of MK2005 indicates that the
results will not be very sensitive to these simplification.  We also
assume that the secondary is ``dark,'' so that the photocenter tracks
the motion of the primary. The position ($z$) of the photocenter of
the primary is thus a function of: 1) the position ($z_0$) at time
$t=0$, 2) the proper motion of the barycenter ($\mu_{z,B}$), 3) the
semi-major axis of the orbit of the primary ($a_o$), 4) the orbital
period and phase $\phi$, and 5) the distance ($d_{pc}$) in pc. (We use
$z$ as shorthand for either $x$ or $y$). With the period in years, the
total mass ($M_{tot}$) in M$_\odot$ and the mass of the companion
($M_{C,J}$) in M$_J$, we find:
\vspace*{-2.0em}
\begin{eqnarray}
  x(t) &=& x_{t=0} + \mu_{x,B}  t + X_o(t)
     \label{eqn:x_t_o} \hspace{1cm}
  y(t) \, =\,  y_{t=0} + \mu_{y,B}  t + Y_o(t)
     \label{eqn:y_t_o}\\
  X_o(t) &=&  a_o \, \cos{( 2 \pi t / P + \phi)}
     \label{eqn:mu_x_o}\hspace{1.1cm}
  Y_o(t) \, = \,  a_o \, \sin{( 2 \pi t / P + \phi)}
     \label{eqn:mu_y_o}\\
  a_o  &=& \frac{0.9547}{d_{pc}} 
     \left( \frac{P}{M_{tot}} \right)^{2/3} M_{C,J} \, ,
     \hspace{5em} [\mas] \label{eqn:a_o_M_J}\\
  \tilde{X}_o(t) &=&                                  a_{o,c\phi} \,      
     -             \left( \frac{2\pi}{P} \right)\; \; a_{o,s\phi} \,  t
     - \frac{1}{2} \left( \frac{2\pi}{P} \right)^2 \, a_{o,c\phi} \,  t^2
     + \frac{1}{6} \left( \frac{2\pi}{P} \right)^3 \, a_{o,s\phi} \,  t^3
        \label{eqn:Xt_t}\\
  \tilde{Y}_o(t) &=&                                  a_{o,s\phi} \, 
     +             \left( \frac{2\pi}{P} \right)\; \; a_{o,c\phi} \,  t
     - \frac{1}{2} \left( \frac{2\pi}{P} \right)^2 \, a_{o,s\phi} \,  t^2
     - \frac{1}{6} \left( \frac{2\pi}{P} \right)^3 \, a_{o,c\phi} \,  t^3
        \label{eqn:Yt_t}
\end{eqnarray}
with $a_{o,c\phi}\equiv a_o \cos{(\phi)}$ and $a_{o,s\phi}\equiv a_o
\sin{(\phi)}$, and where we expand the position change $Z_o(t)$ due to
the orbit to third order to arrive at eqns.~(\ref{eqn:Xt_t}) and
(\ref{eqn:Yt_t}). We identify the orbit-induced position ($z_o$), the
proper motion, acceleration and jerk as the coefficients of the
$t$-terms with powers 0,1,2 and 3, respectively.  $\tilde{X}$ and
$\tilde{Y}$ are a 3\rtp{rd}-order, orbit-based astrometric model. On
the other hand, the observed trajectory can be fit by a polynomial up
to n\rtp{th} order:
\vspace*{-0.7em}
\begin{eqnarray}
z_{F,SIM}(t) &\approx& z_{0,F,SIM} + z_{1,F,SIM} \, t   + z_{2,F,SIM} \, t^2 +
                       z_{3,F,SIM} \, t^3 + {\cal O}(t^4) \, \, ,
   \label{eqn:z_F}
\end{eqnarray}
where the subscript ``$F,SIM$'' indicates that the fit is performed to
the \SIM\ data only. $z_{F,SIM}$ can be evaluated at any previous
epoch and compared with the the observed position at that epoch.  The
position error [$\delta_z(t)$] on $z_{F,SIM}$ depends on the accuracy
of the fit and strongly on the epoch difference.  The difference
between the true position at the \HIPPARCOS\ epoch and the \SIM\
prediction is given by
$\Delta_z(\tau_H) = z_H - z_{F,SIM}(\tau_H)$,
while the significance of $\Delta_z(\tau_H)$ is readily computed. In
order to make a significant detection, $\Delta_z(\tau_H)$ has to be
smaller than the errors on both the \SIM\ prediction and the
\HIPPARCOS\ position.\\
\hspace*{1.4em}
While $z_{F,SIM}$ can fit the space motion during the \SIM\ observing
span extremely well, the {\em extrapolation} of the model to the
\HIPPARCOS\ epoch can lead to large $\Delta_z$ values when the primary
has a companion. In general, large $\Delta_z(\tau_H)$ values indicate
heavy companions, while small values indicate either no companions at
all or a low-mass companion.

\vspace*{-2.5em}
\subsection{Period and Mass Estimates from $\Delta_z(\tau_H)$}
 \label{sec:Period_and_Mass_Estimates}
\vspace*{-0.9em}
Ideally, one would like to know the motion of the barycenter so that
it could be subtracted from $z_{F,SIM}(t)$ to yield the orbital
contribution. In that case, the period follows from the ratio of the
coefficients of eqns.~(\ref{eqn:Xt_t}) and (\ref{eqn:Yt_t}), and would
be independent of orbital phase and inclination. Unfortunately,
because we do not know $\mu_B$, this method can not be used.\\
\hspace*{1.4em}
Alternatively, it is possible to eliminate the phase effects is by
combining the $x$ and $y$ positions differences, at least for face-on
circular orbits. Initial investigations indicate that the effects of
inclination ($i$) are not all that large, as long as $i \la
45^o$. Keeping in mind that the results are indicative rather than
definite, we proceed with circular, face-on orbits. The position
differences can be found analytically \citep{SOSAs}, and read:
\vspace*{-0.6em}
\begin{eqnarray}
   \hspace*{-0.9em}
   \Delta_x \hspace*{-0.7em} &=& 
      \hspace*{-0.7em} x(\tau_H) - \tilde{X}(\tau_H)
      \label{eqn:Delta_x}\hspace*{1cm}
   \Delta_y \, = \, y(\tau_H) - \tilde{Y}(\tau_H)
      \label{eqn:Delta_y}\hspace*{1cm}
   \Delta_{xy} \, = \, \sqrt{ \Delta_x^2 + \Delta_y^2 }
      \label{eqn:Delta_xy}\\
   \hspace*{-0.9em}
   \Delta_{xy,\mu}^2 \hspace*{-0.7em} &=& \hspace*{-0.7em} 
      \frac{a_o^2}{p^2} \left[
      1 
      - 2      s_\alpha  p
      + 2 (1 - c_\alpha) p^2
   \right] 
      \label{eqn:Delta_xy_mu}\hspace*{0.25cm} \& \hspace*{0.25cm}
   \Delta_{xy,\mu+\dot{\mu}}^2 =
      \frac{a_o^2}{p^4} \left[
         \frac{1}{4} 
      +         c_\alpha  p^2
      -  2      s_\alpha  p^3
      +  2 (1 - c_\alpha) p^4
   \right]
      \label{eqn:Delta_xy_mud}
\end{eqnarray}
with $\alpha \equiv 2 \pi \tau_H$, $s_\alpha \equiv \sin{(\alpha/P)}$,
$c_\alpha \equiv \cos{(\alpha/P)}$, and $p = P /(2\pi\tau_H)$, and where
the ``$\mu+\dots$'' subscripts indicate that an expansion of the
orbital motion is used that includes all listed components.  The
position differences can be ratioed to yield a period estimator:
\vspace*{-0.6em}
\begin{eqnarray}
   \tilde{P}_{\mu,\dot{\mu}} &=&\;
      \pi \,\, \tau_H \; \frac{ \Delta_{xy,\mu} }{ \Delta_{xy,\mu+\dot{\mu}} } 
      \; \;\;\;\;  \approx \;
      \left\{
         \begin{array}{cl}
            \ph{\frac{3}{2}} P & \;\;\;\;\; P \; \ll \;    ~  \tau_H\\
                \frac{3}{2}  P & \;\;\;\;\; P \; \ga \;\;  2  \tau_H
         \end{array}
      \right.
      \label{eqn:P_mu_dmu}
\end{eqnarray}
which is accurate for either short or long periods. In the
intermediate regime, $\tilde{P}$ oscillates due to the trigonometric
terms in eqns.~(\ref{eqn:Delta_xy_mud}).  Once the period has been
estimated, the companion mass follows from solving either of the
$\Delta_{xy}$ relations for $a_o$ [and hence mass via
eqn.~(\ref{eqn:a_o_M_J})].\\
\hspace*{1.4em}
The proper motion of the barycenter is not important for this method
because it does not matter how the observed proper motion is divided
between the center-of-mass- and orbital components. The \SIM\ model is
good because it predicts the observed {\em positions} at the \SIM\
epoch, while any position difference at the \HIPPARCOS\ epoch depends
{\em only} on the orbital parameters and $\tau_H$, so that
$\Delta_{xy}$ can be calculated employing the orbital parameters only.

We have performed extensive numerical simulations to test analytical
relations for the position differences derived above
\citep{SOSAs}. Our modeling of the system comprises an implementation
of equations~(\ref{eqn:x_t_o}) with an arbitrary barycentric motion
and a periodic signal in both coordinates (with random phases). We use
this model to predict the position at the \HIPPARCOS\ epoch.  We then
generate, in Monte-Carlo fashion, many random realizations of the
model which are fitted by a polynomial {\em to the SIM positions
only}. The so-determined \SIM\ astrometric model is extrapolated to
the \HIPPARCOS\ epoch to yield $\Delta_{xy}$. We perform: 1) a first
order fit to compute $\Delta_{xy,\mu}$, and 2) a second-order fit for
$\Delta_{xy,\mu+\dot{\mu}}$.  The numerical results are virtually
identical to our analytical predictions
[eqn.~(\ref{eqn:Delta_xy_mud})].

\vspace*{-2.5em}
\section{Results}
 \label{sec:Numerical_Results}
\begin{figure}[!h]
   \vspace{-4em}
   \includegraphics[width=150mm,height=105mm]
   {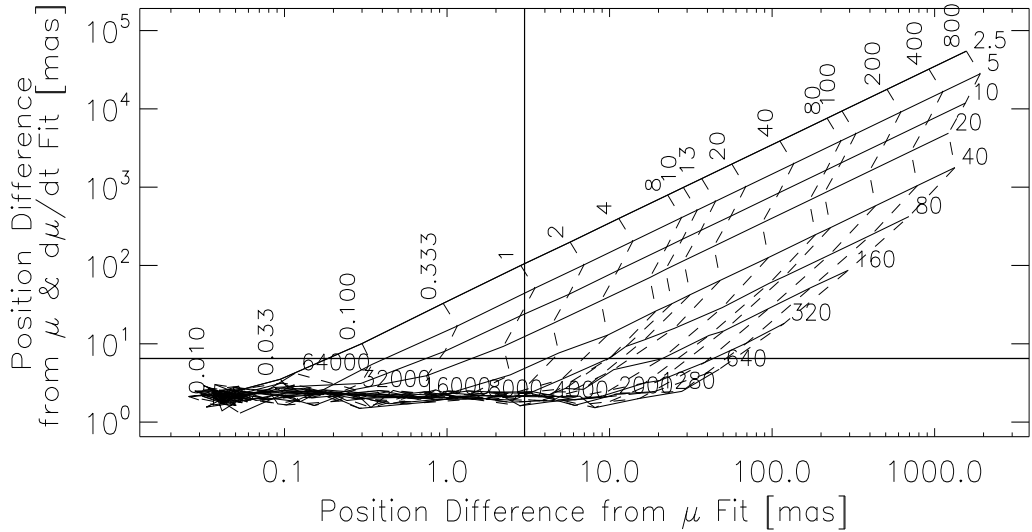}
   \ssROv{0.95}	
   \vspace{-3em} \caption{ 
   \label{fig:SQL_position_residuals} For a series of models with
   various companion masses (vertically oriented numbers in the plot
   in units of $M_J$) and orbital periods (in years; horizontal
   numbers) we plot two positions difference computed from our model
   data. The abscissa is the position difference ($\Delta_{xy,\mu}$),
   while ordinate is $\Delta_{xy,\mu+\dot{\mu}}$. The 3-sigma
   observational limits are indicated by the thick vertical and
   horizontal lines. 
} \esROv
\vspace*{-0.5em}
\end{figure}

\vspace*{-0.9em}
We ran simulations that might be relevant for the \SIM\ quick-look
survey (\S\ref{sec:A_SIM_Quick_Look_Survey_for_HOSAs} below): 5
observations per coordinate per star during a period of 18
months. Conservatively, we assume \SIM\ position errors of 10 \muas\
per observation.  The results are presented in
Figure~\ref{fig:SQL_position_residuals} where we plot two
$\Delta_{xy}$ metrics as determined from the first-order fit
(abscissa) and the second-order fit (ordinate). In this figure,
systems with a given period but with varying mass fall along diagonal
lines from the lower-left to the upper-right (drawn lines)\footnote{
The well-known period-mass degeneracy would result if we were to plot
the fitted $\dot{\mu}$ instead of $\Delta_{xy,\mu+\dot{\mu}}$.}.
In Figure~\ref{fig:SQL_position_residuals}, constant-mass systems are
indicated by the dashed lines. The ``bunching up'' of the lines around
$\Delta_{xy,\mu+\dot{\mu}}=$ 5 \mas\ is due to the limited accuracy of
the \SIM-based determination of the acceleration, and this limit is
used to generate the thick horizontal line (at 3 times this
level). Thus the \SIM\ measurements are the limiting factor for the
determination of $\Delta_{xy,\mu+\dot{\mu}}$ rather than the
\HIPPARCOS\ error. This is the reason why the much less accurate
\GAIA\ data would not be very useful for this application.
Figure~\ref{fig:SQL_position_residuals} indicates that the companion's
mass and period can be determined in a large region of parameter
space. The residuals suggest that the orbits of a 1 (10) $M_J$ ESGPs
can be characterized up to periods of 10 (80) years, while this is
possible for stellar companions up to 500 years\footnote{
The 1\rtp{st}-order $\Delta_{xy,\mu}$ values are significant up to
1,000 years at the Hydrogen burning limit and 4,000 years for a double
star with solar-mass components.}. If we use the expected \SIM\
accuracy of 4 \muas, this period range is extended by a factor 2 -- 4.

\vspace*{-2.5em}
  \section{A SIM Quick-Look Survey for HOSAs}
\label{sec:A_SIM_Quick_Look_Survey_for_HOSAs}
\vspace*{-0.9em}
Figure~\ref{fig:SQL_position_residuals} also indicates that a few
highly accurate observations suffice to identify planetary, MS and BD
companions. Such data could be generated by a SIM quick-look (SQL)
survey. The targets are bright \HIPPARCOS\ stars, so we assume that an
SQL observation can be achieved in one minute per position per
baseline. Thus, one-thousand stars can be done in 10,000 minutes (6.9
days), so that several thousand stars can be included in an SQL survey
without impacting the overall \SIM\ mission significantly. Given that
the PDF for ESGPs predicts heavy solar-system analogs around 7.9\% of
stars, a survey of $\sim$5,000 stars may find 400 HOSAs. Such a sample
would firmly establish the PDF in the long-period regime and indicate
how unique the Solar system really is.\\
\hspace*{1.4em}
In order to maximize the yield of HOSAs (and some SOSAs), an SQL
program needs to avoid MS and BD multiples. The subset of 73,000
\ARIHIP\ stars \citep{ARIHIP} that show no signs of binarity is a good
starting point for the target selection of an SQL survey. \\
\hspace*{1.4em}
Those systems that do not show signs of binarity in the
SQL+\HIPPARCOS\ survey are likely to have either sub-stellar
companions or stellar companions with very long-periods. Those systems
warrant further SIM observations.  The SQL follow-up survey of those
stars with suspected sub-stellar companions would be significantly
more sensitive than the SQL survey. Figures similar to
figure~\ref{fig:SQL_position_residuals} but with employing the SQL
follow-up data indicate (not shown) that the ESGPs can be detected
with masses as low as 0.1 M$_J$ in 10 year orbits. Period estimation
for 1 [10] M$_J$ is extended by a factor four [two] (to 40 [160]
years).

\vspace*{-2.5em}
  \section{Conclusions}
\label{sec:Conclusions}
\vspace*{-0.9em}

A judicial combination of \HIPPARCOS\ data, a SIM quick-look survey
and follow-up SIM observations at full accuracy can uncover several
hundred extra-solar planetary systems with periods comparable to the
gas giants of the Solar system. Such a program is only possible with
\SIM-like accuracies, and the results would nicely complement radial
velocity and imaging surveys.\\
\hspace*{1.4em}
Given the importance of accurate pre-\SIM\ astrometry, it is sad to
realize that the canceled {\em FAME} mission \citep{FAME2003} would
have provided an excellent reference catalog for detection and
characterization of solar-system analogs. Likewise, it is of
pre-eminent importance to continue all-sky astrometric programs at
intervals of ten to twenty years to probe the long-period regime. The
required accuracy depends on the desired period- and mass ranges, but
a survey with an accuracy at the \HIPPARCOS\ level (one-half to one
$\mas$) would already be very valuable.

\snROv{0.90}

\begin{flushleft}

\end{flushleft}

\enROv

\end{document}